\documentstyle[12pt]{article}

\def\onlinecite#1{\cite{#1}}
\def\ds{\displaystyle}

\begin{document}
\begin{flushright}
BU-HEP/92-17 \\
TAUP 2034-93\\
hep-ph/9304217\\
April 1993
\end{flushright}
\vskip1cm
\begin{center}
{\LARGE  STATIC PROPERTIES OF QUARK SOLITONS}
\vskip1cm
{\large Gonen Gomelski$^{*}$
 and Marek Karliner$^\dagger$}
\vskip0.7cm
{\it School of Physics and Astronomy
\\ Raymond and Beverly Sackler Faculty of Exact Sciences
\\ Tel-Aviv University, 69978 Tel-Aviv, Israel }
\vskip0.7cm
{\large Stephen B. Selipsky$^{\hbox{\scriptsize \P}}$}
\vskip0.7cm
{\it Department of Physics, Boston University\\
Boston, MA  02215-2507, U.S.A.}
\end{center}
\vskip 1.5cm
\begin{abstract}
It has been conjectured that at distances smaller than the confinement
scale but large enough to allow for nonperturbative effects, QCD is
described by an effective $SU(N_c {\times} N_f)_L\times SU(N_c
{\times} N_f)_R$ chiral Lagrangian.  The soliton solutions of such a
Lagrangian are extended objects with spin ${1\over 2}$.  For
$N_c{=}3$, $N_f{=}3$ they are triplets of color and flavor and have
baryon number ${1\over3}$, to be identified as constituent quarks.  We
investigate in detail the static properties of such constituent-quark
solitons for the simplest case $N_f{=}1, N_c{=}3$.  The mass of these
objects comes from the energy of the static soliton and from quantum
effects, described semiclassically by rotation of collective
coordinates around the classical solution.  The quantum corrections
tend to be large, but can be controlled by exploring the Lagrangian's
parameter space so as to maximize the inertia tensor.  We comment on
the acceptable parameter space and discuss the model's further
predictive power.
\end{abstract}

\newpage
\section{Introduction}
One of the outstanding questions in particle physics continues to be
the relationship between the phen\-o\-men\-ologically successful
non-relativistic con\-stit\-uent-quark model of hadrons, and QCD's
fundamental degrees of
freedom\cite{GM}-\cite{Bjorken}.
The $u,\ d,\ s$
valence quarks appearing in hadron spectroscopy have the same quantum
numbers as the QCD current-quark fields, yet are much heavier and in
some sense embody the strongly interacting QCD dynamics.  The
experimental results\cite{EMC} on spin structure of the proton
indicate that the constituent quarks ought to be thought of as
composite objects with internal structure.  Finding a theoretical
description of this structure is an important and interesting
challenge.

Kaplan has put forward a model\cite{Kaplan} combining some of the
features of the chiral quark\cite{GM} and
skyrmion\cite{Skyrme,SM,WittenLewes} approaches.  In this model it is
postulated that at distances smaller than the confinement scale but
large enough to allow for nonperturbative phenomena the effective
dynamics of QCD is described by a chiral Lagrangian whose target space
is \break
$SU(N_c {\times} N_f)_L\times SU(N_c {\times} N_f)_R $, resulting
 from misalignment of the $\bar {q}q$ condensate in the color space.
One then makes the additional assumption that the solitons of this
effective theory are stable and may be quantized semiclassically.
Remarkably enough, it turns out that such solitons have the same
quantum numbers as the constituent quarks in the quark model.  Thus
the constituent quarks in this model are ``skyrmions'' in color space.

This is a very attractive idea, but some crucial elements of the
puzzle are still missing, most importantly a derivation of the
effective dynamics from QCD.  However, recent work on QCD in 1+1
dimensions\cite{QCDtwo} (QCD$_2$) provides evidence in support of this
picture.  Exact nonabelian bosonization allows the Lagrangian of
QCD$_2$ to be re-expressed in terms of a chiral field $u(x) \in
U(N_c{\times}N_f)$ and of the usual gauge field.  This bosonized
Lagrangian has topologically nontrivial static solutions that have the
quantum numbers of baryons and mesons, constructed out of constituent
quark and anti-quark solitons.  The identification of these solitons
with constituent, rather than current quarks, goes beyond the usual
fermion-boson duality in two dimensions.  It is based on the fact that
their mass is generated dynamically and depends on the coupling
constant.

When $N_c$ solitons of this type are combined, a static,
finite-energy, color singlet solution is formed, corresponding to a
baryon.  Similarly, static meson solutions are formed out of a soliton
and an anti-soliton of different flavors.  The stability of the mesons
against annihilation is ensured by flavor conservation.  These results
can be viewed as a derivation of the constituent quark model in
QCD$_2$.  Thus the idea of constituent quarks as solitons of a
Lagrangian with colored chiral fields becomes exact in
\hbox{$D = 1{+}1$}.

Pending a more formal justification of the effective dynamics in
\hbox{$D = 3{+}1$}, it is interesting to see whether the quark
solitons or ``qualitons'' have phenomenologically acceptable static
properties, beyond their quantum numbers.  The model discussed in
Ref.~\onlinecite{Kaplan} is the simplest one with the required
properties.  It is the analogue of the Skyrme model in color space.

It is quite probable that the actual effective Lagrangian in four
dimensions is considerably more complicated, but retains the same
transformation properties under flavor and color.  Past experience
with skyrmion-like models shows that their qualitat\-ive predictions are
remarkably independent of specific details of the
Lagrangian\cite{ModelIndep,extraterms}.  The group theoretical analysis
of collective-coordinate quantization\cite{Kaplan} is thus likely to
be correct, regardless of the specific dynamics.  So until the correct
approximate dynamics is found, it is worthwhile to investigate in some
detail the simplest model possessing the right symmetries.

\section{Model and Solution}
We start from the Lagrangian\cite{Kaplan}
\begin{eqnarray}
{\cal L}={-}{1\over4g^2_s} G^a_{\mu\nu}G^{a\mu\nu} {+}
{1\over4}\mu_\chi^2f^2 \bigg\{ {-}{\rm Tr}\hat R_\mu \hat R^\mu
+{1\over4}c_4{\rm Tr}\left[\hat R_\mu,\hat R_\nu\right]^2
+c_\alpha {\alpha_s\over4\pi}{\rm Tr}T_a\Sigma T_a\Sigma^\dagger
{+}\cdots\bigg\}
\label{Lagrangian}\end{eqnarray}
plus a Wess-Zumino term, where $G^a_{\mu\nu}$ is the gluon
field strength,
$\mu_\chi$ is the chiral symmetry breaking scale,
$f$ is analogous to the pion decay constant $f_\pi$,
$T_a$ is a color $SU(3)$ generator normalized by
$\hbox{Tr} T_a T_b = {\scriptstyle 1\over 2}\delta_{ab}$,
${\hat R}_\mu = \left( \Sigma^\dagger D_\mu\Sigma\right)/\mu_\chi $,
$\Sigma =\exp(2i\Pi^aT_a/f),\ D_\mu= \partial_\mu+iA^a_\mu T_a$
is the color covariant derivative,
$\alpha_s = g_s^2/4\pi$
is the effective color fine structure constant and the dots ($\cdots$)
refer to other operators with four or more derivatives and higher
powers of $\alpha_s$.

Of the five parameters in (\ref{Lagrangian}) we have some experimental
information only on $f$ and $\alpha_s$, and even that is quite
limited.  The two-flavor model's axial flavor current
suggests\cite{Kaplan} that to leading order in $\alpha_s$, $f \approx
f_\pi/{\sqrt3} \approx 53.7 \hbox{\ MeV}$, but higher derivative
operators ignored in this treatment might renormalize $f$ away from
that value, just as in the usual Skyrme model\cite{SM}.  Here
$\alpha_s$ is the strength of the effective color interaction between
the color-carrying condensate and the gluon field, in principle
derivable by nonperturbatively matching the effective and fundamental
theories\cite{GM}.  As no such calculation is available at present, in
practice we can take as a rough estimate the value $\alpha_s \sim
0.28$ provided by the non-relativistic quark model.  For the other
parameters, dimensional analysis suggests\cite{GM,Kaplan} that
$\mu_\chi \sim 4\pi f$ and $c_4 \sim 1 \sim c_\alpha$.

Given a parameter set $\{\alpha_s,c_4,c_\alpha,f,\mu_\chi\}$, we
seek classical static solutions for both $\Sigma$ and gauge fields,
adopting a radially-symmetric skyrmion ansatz
\begin{equation}
\Sigma = e^{iF(r){\bf \hat r}\cdot{\vec\sigma}_c},
\quad F(0)=\pi\;,\quad F(\infty) = 0\;, \label{sigmaansatz}
\end{equation}
where ${\vec\sigma}_{\rm c}$ acts on a two-dimensional subspace of
$SU(3)_c$.  To obtain the classical gluon configuration we note that
it should have the same symmetry as $\Sigma$ under combined spin and
color isospin rotations.  We work in $A_0=0$ gauge and eliminate the
remaining gauge freedom, imposing the Gauss law constraint for unit
winding number.  We also note that energy minimization will remove the
pure Yang-Mills terms, resulting in a configuration whose nonvanishing
components are
\begin{eqnarray}
A^i_m    & =&{\gamma (r)\over r}
             \epsilon_{imj}\hat r_j\;,\qquad m = 1,2,3\;,
\nonumber\\
{\bf\vec A_8}& =& -t{g_s^2\over\sqrt{12}}{{\bf\hat r}\over 8\pi^2 r^2}
           \left[2(\pi{-}F) + (1{+}\gamma)^2\sin{2F}\right] \;.
\label{gluonansatz}\end{eqnarray}
Then the radial energy density of a qualiton
(relative to the classical vacuum) is
\begin{eqnarray}
 \rho(r) &=& {\alpha_s \over 96\pi^2} {1\over r^2}
           \left[2(\pi - F)+(1+\gamma)^2 \sin{(2F)} \right]^2
\nonumber\\
 &+&{1\over 2r^2 \alpha_s}\left[2(r\gamma')^2 + \gamma^2(2+\gamma)^2\right]
\nonumber\\
 &+&2\pi f^2\left[(rF')^2+2(1+\gamma)^2 \sin^2{F}\right]
\nonumber\\
 &+&{c_4\over r^2}{{4\pi f^2}\over \mu_\chi^2}\sin^2{F} (1+\gamma)^2
   \left[2(rF')^2+(1+\gamma)^2 \sin^2{F}\right]
\nonumber\\
 &+&c_\alpha{\alpha_s\over 2}\mu_\chi^2 f^2 r^2
   \left[2-\cos{F}-\cos^2{F}\right]
\;,\label{equaliton}\end{eqnarray}
and the classical contribution to the rest mass is
\begin{equation}
M_{cl}=\int_0^{\infty}{\rm d}r \rho(r) \;. \label{totmas}
\end{equation}
The five terms represent the respective contributions of the color
electric field, the color magnetic field, the kinetic term, the
four-derivative term, and the symmetry-breaking term.  Minimizing the
classical mass yields the variational equations for $F(r)$ and
$\gamma(r)$:
\begin{eqnarray}
&
4\pi f^2\left[r^2{+}
{\ds{4c_4\over\mu_\chi^2}}\sin^2{F}(\gamma{+}1)^2\right]F''
- 4\pi f^2\bigg[(\gamma{+}1)^2\sin{2F}{-}2rF'\bigg]
&
\nonumber\\
&
- {\ds{\alpha_s\over 24\pi^2 r^2}}
 \bigg[2(\pi{-}F){+}(\gamma{+}1)^2 \sin{2F}\bigg]
 \bigg[(\gamma{+}1)^2 \cos{2F}{-}1\bigg]
&
\nonumber\\
&
- {\ds{8c_4\pi f^2\over\mu_\chi^2}}
 \bigg[{\ds{1\over r^2}}\sin{2F}\sin^2{F} (\gamma{+}1)^4{-}
 \sin{2F}(\gamma{+}1)^2(F')^2 {-} 4\sin^2{F}(\gamma+1)\gamma'F'\bigg]
&
\nonumber\\
&
-{1\over 2}c_\alpha\alpha_s\mu_\chi^2 f^2r^2(\sin{F}+\sin{2F})\,=\,0\;;
\label{vareq1}
\end{eqnarray}
\begin{eqnarray}
&
{\ds{1\over\alpha_s}}\gamma''
- {\ds{1\over\alpha_s r^2}}\gamma(\gamma{+}1)(\gamma{+}2)
- 4\pi f^2(\gamma{+}1)\sin^2{F}
&
\nonumber\\
&
- {\ds{\alpha_s\over 48\pi^2 r^2}}(\gamma{+}1)\sin{2F}
 \bigg[2(\pi{-}F){+}(\gamma{+}1)^2\sin{2F}\bigg]
&
\nonumber\\
&
- {\ds{8c_4\pi f^2\over\mu_\chi^2}}
\left[(\gamma{+}1)\sin^2{F}(F')^2+(\gamma{+}1)^3
  \,{\ds{\sin^4{F}\over r^2}}\right]\,=\,0
&
\label{vareq2}
\end{eqnarray}
subject to the boundary conditions (\ref{sigmaansatz}) and
$\gamma(0)\,=\,0\,=\,\gamma(\infty)$.
The equations were solved with the COLSYS package\cite{COLSYS}.

Given a particular solution $\{\Sigma, A\}$, minimizing the classical
mass (\ref{totmas}), the symmetry generates a family of new solutions
with the same energy:
\begin{eqnarray}
\Sigma \rightarrow \Omega\,\Sigma\,\Omega^\dagger\ ,\qquad
     A \rightarrow \Omega\, A\, \Omega^\dagger
     +i\Omega\nabla\Omega^\dagger \;, \label{eqA}
\end{eqnarray}
where $\Omega({\bf\vec r},t) \in SU(3)$ is a collective coordinate.
Quantizing these zero modes in the gauged case\cite{Kaplan} is more
subtle than in the global-symmetry Skyrme model because the quantum
states have to satisfy the Gauss law constraint.  In the external
background $\{\Sigma, A\}$ corresponding to a given parameter set this
constraint leads to differential equations for $\Omega$.  We solve
these equations numerically, obtaining the two moments of inertia
$I_1$ and $I_2$, which govern the contribution from quantization of
the collective coordinates in the full semiclassical mass formula
\begin{equation}
M=M_{cl}+j(j+1)\left({1\over2I_1}-{1\over2I_2}\right)+{1\over2I_2}
\left(C_2-{1\over{12}}\right) \;, \label{eqB}
\end{equation}
where $C_2$ and $j$ are respectively the $SU(3)$ Casimir and the spin
in the appropriate representation.  The ground state solitons are
color triplets with $j={1\over 2}$, and mass
\begin{equation}
M=M_{cl}+{3\over8I_1}+{1\over4I_2}
\equiv M_{cl}+M_1+M_2\;. \label{totm}
\end{equation}
In the analogous expression in the usual Skyrme model the quantum
contribution, $M_1+M_2$, is suppressed by $1/N_c^2$ relative to the
classical mass.  This justifies the semiclassical approach in the
Skyrme model.  In the present case, however, the large-$N_c$ expansion
does not help, because both the classical mass and the moments of
inertia scale like $\sim N_c^0$: parametrically the classical and
quantum terms are of the same order.  For the semiclassical approach
to make sense, the parameters of the model have to be chosen so as to
make the quantum contribution numerically small, while at the same
time giving phenomenologically acceptable results.

Starting with the ``natural'' guess discussed earlier,
$\{\alpha_s = 0.28$, $c_4 = 1$, $c_\alpha = 1$, $f = 53.7$ MeV,
$\mu_\chi = 4\pi f\}$, we obtain
$M = M_{cl}{+}M_1{+}M_2 = (361{+}605{+}443) = 1409$ MeV.
This is doubly unacceptable: $M$ is four times the phenomenological
value of 350 MeV, and the ratio of quantum corrections to classical
contribution is large,
\hbox{$\eta \equiv (M_1{+}M_2)/M_{cl} = 2.2$.}
We can also compute the ``half-height radius'' $r_{1\over2}$ defined
by $F(r_{1\over2}) = \pi/2$, and obtain $r_{1\over2} = 0.354$ fm,
which does look reasonable compared to the proton charge radius
$\approx0.86$ fm.  This situation is rather generic: it is easy to
find parameters yielding a smaller mass with a bigger radius or vice
versa, but one cannot keep both of them small, while decreasing
$\eta$: moments of inertia scale like $M_{cl} r^2$ and so from
(\ref{totm}) $\eta \sim 1/(M_{cl}r)^2$.

We can learn more by systematically investigating the parameter space.
This task can be simplified by expressing all dimensionful parameters
in terms of $f$.  Then the \lower-0.5ex\hbox{$\chi$}SB scale $\mu_\chi$
appears only in the dimensionless combinations ${\tilde c_4} \equiv
(4\pi f/\mu_\chi)^2 c_4$ and ${\tilde c_\alpha} \equiv (\mu_\chi/4\pi
f)^2 c_\alpha$.  Together with $\alpha_s$ they form a three
dimensional parameter space, which determines the physics up to an
overall scale factor.  A variational estimate then suggests that $M$
has a local minimum, $\sim 20f$, near ${\tilde c_4} \sim 3$.  In the
${\tilde c_\alpha}$ direction the minimum occurs only at the boundary,
${\tilde c_\alpha} = 0$.  Near the minimum the mass changes slowly as
a function of ${\tilde c_4}$, allowing partial minimization of either
$\eta \sim 2/({\tilde c_4})^{2}$, or of $r_{1\over2}f \sim
0.1\sqrt{\tilde c_4}$.

Numerical results confirm these expectations.  With $\alpha_s = 0.28$
and ${\tilde c_\alpha} = 1$, a local minimum $M = 18.3f$ occurs at
${\tilde c_4} = 2.72$, yielding $\eta = 0.377$.  When $c_\alpha$
is fine-tuned small, the minimum mass becomes $M\approx 16f$, falling
to $M\approx 13f$ for $\alpha_s = 1$.  Thus for any reasonable values
of $c_\alpha$ and $\alpha_s$, $M$ is well above the colored
pseudo-Goldstone mass
$m_\Pi = ({3\over2}{\tilde c_\alpha}\,4\pi\alpha_s)^{1/2} f$, and
comparable to $\mu_\chi\sim 4\pi f$.  Furthermore, a 350 MeV
constituent quark requires $f$ to be considerably below 53.7 MeV,
implying a large value for $r_{1\over2}\approx 0.35\sqrt{\tilde c_4}\,
(53.7\, {\rm MeV}/f)$ fm.  Clearly, an equally valid alternative is to
take a larger $f$, so as to get a smaller $r_{1\over2}$, at the
expense of increasing $M$ above 350 MeV.  For example, $f=36$ MeV
with $c_\alpha$ small yields $r_{1\over2}=0.86$ fm, $M=570$ MeV,
with $\eta$ unchanged.

At first it therefore seems that to obtain realistic properties of
constituent quarks one must modify the model--- which, as noted
earlier, is only the simplest of a large family of Lagrangians
possessing the right symmetries.  The variational argument indicates
that additional four-derivative terms are unlikely to make a
qualitative difference, but it also indicates possibilities for
quantitative improvements.

We would like to point out, however, that there is another, somewhat
unusual possibility: the radius of a constituent quark might actually
be larger than the radius of the proton.  We are used to composite
objects being larger than the building blocks out of which they are
constructed, but there are some exceptions to this rule when the
interactions among the constituents are very strong.  Clearly, when
something like that happens, there must be a very substantial
deformation of the original constituents.  An example is provided by
the fusion of two Deuterium nuclei\cite{Horn} to form a nucleus of
Helium-4.  The r.m.s.\ charge radius of $^2$H is 2.12 fm, with binding
energy of 2.2 MeV, while the corresponding figures for $^4$He are 1.67
fm and 28.3 MeV.  The confining interaction between qualitons is
stronger than the interactions that shape the qualiton.  It is
therefore possible that a similar phenomenon occurs when qualitons
bind together to form hadrons.  We do not know whether Nature realizes
this scenario, or whether the effective Lagrangian then remains a
useful guide, but if it does, we can find a broad region of the
parameter space corresponding to a phenomenologically acceptable mass.

To this effect, we first choose a somewhat small $\tilde c_\alpha$, to
minimize the dimensionless mass.  This will allow $M = 350$ MeV to
correspond to a slightly larger $f$.  Next we decrease $\tilde c_4$ in
the neighborhood of the minimum, so as to decrease the radius, while
constraining $\eta < 0.33$ to protect the validity of the
semi-classical approximation.  With $\alpha_s = 0.28$, $c_\alpha =
0.1$, and $\mu_\chi = 4\pi f$, we thus find parameters $c_4 = 2.46, f
= 21.4$ MeV, giving $r_{1\over2}=1.43$ fm and $\eta=0.33$.  Fig.~1
plots the resulting shape functions.

To test the sensitivity of this choice we note that taking instead
$c_\alpha = 0.3$ decreases $f$ only slightly to 20.6 MeV, and
increases $r_{1\over2}$ to 1.50 fm, with a range \hbox{1.44 fm $ <
r_{1\over2} < 1.56$ fm} corresponding to \hbox{$0.1 < \alpha_s < 1$}.
Even taking $c_\alpha = 1, \alpha_s = 1$ raises $r_{1\over2}$ only to
1.67 fm.  Thus in the phenomenologically relevant regions of parameter
space the results are rather stable and do not require fine-tuning.

\section{Summary and outlook}
We have studied the static properties of constituent quark solitons
for the simplest case $N_f{=}1$, $N_c{=}3$.  Self-consistency of the
semiclassical calculation indicates that in order to obtain
phenomenologically acceptable constituent quark masses, such objects
are likely to have rather large radii, before binding into hadrons.  A
more detailed investigation of this possibility requires extending the
present work to multi-soliton configurations.  Such an extension would
also make it possible to compute modifications of constituent quark
color magnetic moments in bound states, giving baryon hyperfine mass
splittings and thus fixing the effective $\alpha_s$.

It would also be interesting to extend the calculation to an
$N_f{\ge}2$ model, since for $N_f{=}1$ there is no chiral flavor
symmetry and hence no pion.  This would allow a computation of the
constituent quark axial-vector coupling $g_A^q$.  The framework can
also be fruitfully applied in the study of the spin and strangeness
content of the proton, or in a computation of the Isgur-Wise
function\cite{IsgurWise} representing the light-quark degrees of
freedom in the formalism of the heavy quark symmetry.

\vskip 2cm
\begin{flushleft}
{\large\bf Acknowledgements}
\end{flushleft}
We thank David Kaplan and Yitzhak Frishman for useful discussions.

The research of M.K.\ was supported in part by the Basic Research
Foundation administered by the Israel Academy of Sciences and
Humanities and by grant No.\ 90-00342 from the United States-Israel
Binational Science Foundation (BSF), Jerusalem, Israel.

S.B.S.\ acknowledges the support of a Texas National Research
Laboratory Commission SSC Fellowship Award.  This work was also
supported under TNRLC grant RGFY92B6, and under NSF contract
PHY-9057173 and DOE contract DE-FG02-91ER40676.

\vskip 2cm
\begin{flushleft}
{\large\bf Figure Caption}
\end{flushleft}
\noindent The profiles $F(r)$, $\gamma(r)$, and the classical
mass density $\rho(r)$ in MeV/fm, with $\alpha_s = 0.28$,
$c_4 = 2.46$, $c_\alpha = 0.1$, $f=21.4$ MeV, and $\mu_\chi=4\pi f$.
$M = 350$ MeV, $\eta\equiv M_{quantum}/M_{cl} = 0.33$.

\end{document}